\documentclass{revtex4}

\usepackage{graphicx}
\def\TeV{\ifmmode {\,\mathrm{ Te\kern -0.1em V}}\else
                   \textrm{Te\kern -0.1em V}\fi}%
\def\GeV{\ifmmode {\,\mathrm{ Ge\kern -0.1em V}}\else
                   \textrm{Ge\kern -0.1em V}\fi}%
\def\MeV{\ifmmode {\,\mathrm{ Me\kern -0.1em V}}\else
                   \textrm{Me\kern -0.1em V}\fi}%
\def\keV{\ifmmode {\,\mathrm{ ke\kern -0.1em V}}\else
                   \textrm{ke\kern -0.1em V}\fi}%
\def\eV{\ifmmode  {\,\mathrm{ e\kern -0.1em V}}\else
                   \textrm{e\kern -0.1em V}\fi}%

\newcommand{\ee}    {\mathrm{e}^+\mathrm{e}^-}
\newcommand{\mumu}  {\mu^+\mu^-}
\newcommand{\ttb}   {{\mathrm t\bar{\mathrm t}}}
\newcommand{\bb}    {{\mathrm b\bar{\mathrm b}}}
\newcommand{\cc}    {{\rm c\bar{\rm c}}}

\newcommand{\cm}{\mathrm{cm}}

\newcommand{\micron}{\mu\mathrm{m}}

\newcommand{\meuro}{\,{\sc Meuro}}
\begin{document}
\bibliographystyle{unsrt}

% number  and your conference paper identification number on the
% title page in preprint mode. Multiple \preprint commands are allowed.
%\preprint{Snowmass E3029}
\begin{flushright}
Snowmass E3029
\end{flushright}
\title{The TESLA Detector}
% Optional argument for running titles on pages
%\title[]{}



\author{Klaus M\"onig}
\email[]{Klaus.Moenig@desy.de}
%\thanks{}
\affiliation{DESY-Zeuthen}


\date{October 15th 2001}

\begin{abstract}
For the superconducting linear collider TESLA a multi purpose detector
has been designed.
This detector is optimised for the important physics processes expected at a
next generation linear collider up to around 1 TeV and is designed for the
specific environment of a superconducting collider.
\end{abstract}

% \pacs{}

%\maketitle must follow title, authors, abstract and \pacs
\maketitle
\begin{center}
Talk presented at the 2001 Snowmass Workshop on the Future of Particle Physics
July 1-20, 2001
\end{center}

\section{Introduction}
\label{sec:intro}
Recently the Technical Design Report for the superconducting linear collider
TESLA has been completed \cite{tdr_tesla}. As a proof that the proposed
physics program \cite{tdr_phys} can be done with a detector of known
technology and reasonable price the conceptual design of a multipurpose
detector has been included \cite{tdr_det}.

The physics goals of TESLA require from the detector
\begin{itemize}
\item very good momentum resolution ($\delta 1/p \sim 4\cdot 10^{-5}/\GeV$) 
e.g. to measure the Z recoil mass in the process $\ee \rightarrow {\rm ZH},
{\rm Z}\rightarrow \ell^+ \ell^-$,
\item high resolution of the hadronic jet energy 
  ($\Delta E/E \approx 30\%/\sqrt{E}$) to reconstruct multi-jet events with
  intermediate resonances such as ZHH or $\ttb$H,
\item superb b-tagging to identify multi-b final states  like ZHH or $\ttb$H
  or to separate ${\rm H} \rightarrow \bb, \, {\rm H} \rightarrow \cc$
  and ${\rm H} \rightarrow gg$,
\item optimal hermeticity to reduce backgrounds in missing energy channels,
  especially to veto two-photon induced events.
\end{itemize}
At TESLA a bunch train consists of around 2800 bunches with a bunch spacing of 
more than 300 ns. The relatively long time between bunches makes bunch
identification easy and no special fast detectors are needed for this purpose.
The only really relevant background at TESLA are $\ee$ pairs created in the 
collision. These pairs are concentrated at low angle or low transverse momentum.
The low angle component requires a mask in the forward direction, which can,
however, to a large part be used for calorimetry. The low $p_t$ component
at large angles
can be kept at low radius by a strong magnetic field. It limits the radius
of the most inner detector layer to around 1.5\,cm and causes a significant
background in that layer of the vertex detector.

\begin{figure}[htb]
\begin{center}
\includegraphics[width=0.45\linewidth,bb=0 0 506 488,clip]
  {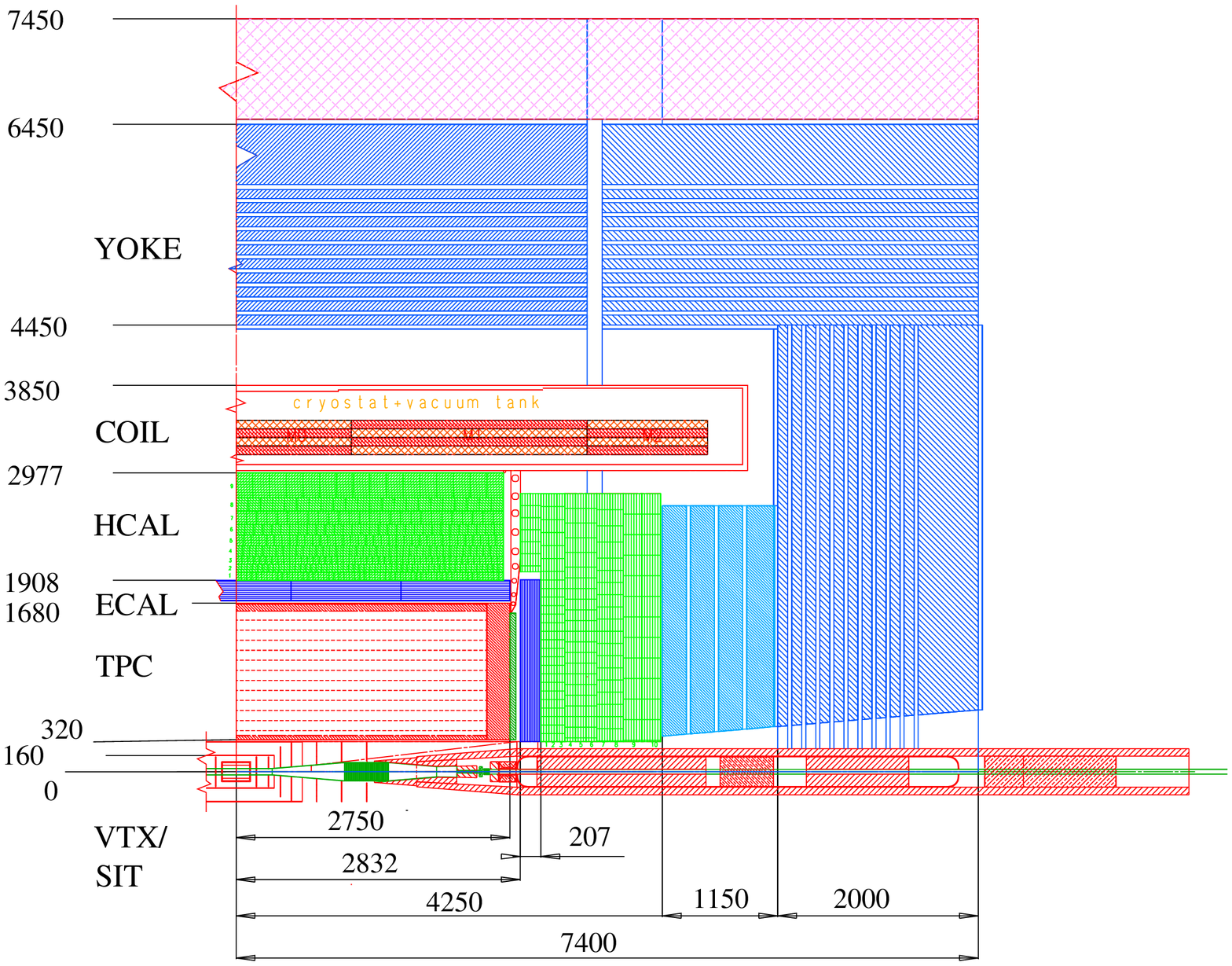}
\hfill
\includegraphics[width=0.45\linewidth]{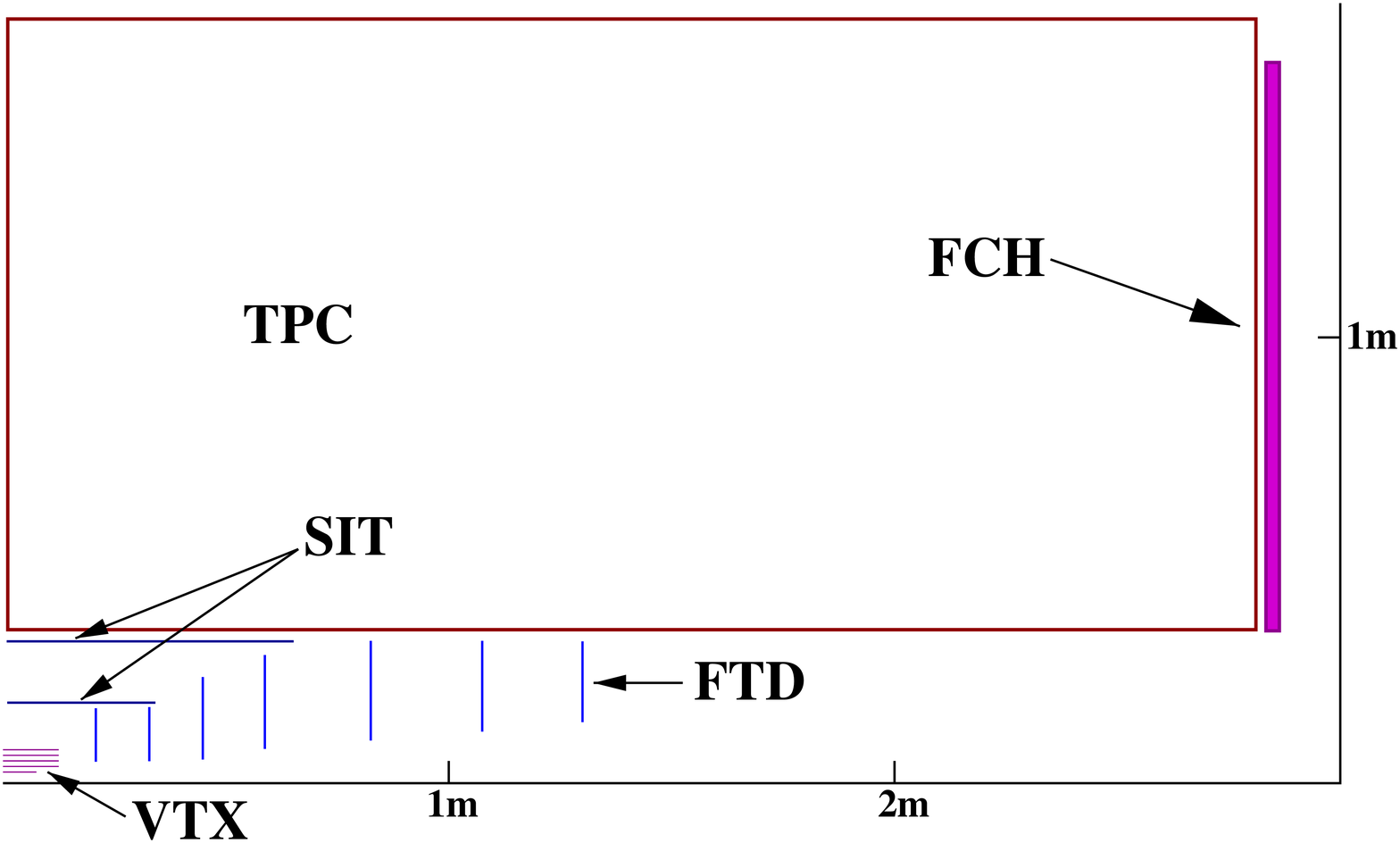}
\end{center}
\caption{a) Layout of the TESLA detector
b) The tracking system of the TESLA detector}
\label{fig:fuldet}
\end{figure}

Figure \ref{fig:fuldet} a) shows the principle layout of the detector. 
The tracking system and the calorimeters are situated inside a 4T 
superconducting coil.
Sections \ref{sec:track} and \ref{sec:calo} describe the two main subsystems, 
tracking and calorimetry. In section \ref{sec:perf} the expected performance
of the detector is summarised. Needed and planned R\&D to build the detector
presented here is described in other talks of this session.

\section{The Tracking system}
\label{sec:track}
The layout of the tracking system is  shown in figure \ref{fig:fuldet} b).
It consists
of a a precise microvertex detector, a large TPC, silicon tracking in between
these two main detectors and a set of forward chambers behind the TPC endplate.

The TPC is in principle similar to existing ones, such as ALEPH, DELPHI or 
STAR.
%A detailed description can be found in \cite{ref:tpc}. 
Due to the higher photon background higher granularity and 100--200 pad rows 
are needed. For this reason an R\&D program has 
been started to read out the signals with new technologies like GEMs or 
Micromegas. Not to compromise the momentum resolution and the calorimetric
energy reconstruction also the field cage and the endplate will be designed to 
be as thin as possible.

For the microvertex detector (VTX) several technologies are under study, CCDs, 
active pixel sensors and CMOS sensors. With the CCDs a point resolution of
$3.5 \micron$ and a layer thickness of $0.12 \%$ of a radiation length
can be reached.
%This would result in an impact parameter resolution of 
%$\sigma_{IP} \sim 2.9 \micron \oplus \frac{3.9 \micron}{p \sin^{3/2}\theta}$.
The main problem of CCD detectors for TESLA is the relatively long
readout time. Reading out the columns in parallel with a frequency of
50\,MHz results in one complete readout every 100 bunch crossings giving
a relatively high but acceptable background in the innermost layer.
For the other technologies an R\&D program is in progress to reach a similar 
performance as presented for the CCDs.

The intermediate tracking detector consists of two cylinders of silicon
strip detectors in the barrel region down to $25^\circ$ (SIT) 
and three silicon pixel and four silicon strip layers on either side in the 
forward region (FTD) of the
detector. The SIT and FTD use only technologies that have already been used 
successfully in LEP so that important R\&D is not needed beyond finding ways 
to thin the detectors. It improves 
significantly the momentum resolution in the full acceptance region
and enables a precise angle measurement in the forward region before a lot
of material is crossed, which is needed to measure the acolinearity of 
Bhabha events in the analysis of the luminosity spectra.

To get a reasonably precise momentum measurement below a polar angle of around
$12^\circ$, where the projected tracklength in the TPC starts to limit the
accuracy, at least one precise
space point with a resolution around $50 \micron$ at maximal distance from 
the interaction point is needed. This can for example be done with a 
forward chamber (FCH) consisting of 12 straw tube planes with $100 \micron$
resolution. The high redundancy helps solving ambiguities in the
pattern recognition. The chamber is also extended over the full TPC endplate
to help mapping distortions in the TPC and possibly to serve as a preshower
device.

\begin{figure}[t]
\begin{center}
\setlength{\unitlength}{0.333cm}
\begin{picture}(18,12)\thicklines
\put(0,0){\mbox{
\includegraphics[height=4.cm,bb=65 0 595 338,clip]{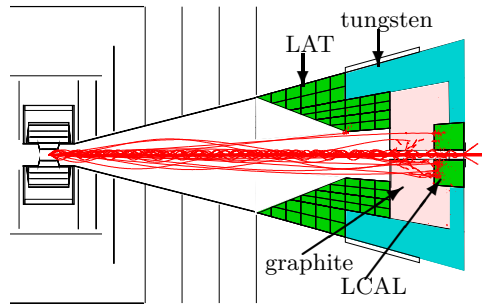}}}
\put(12,10){\vector(0,-1){1.3}}
\put(11.3,10.1){LAT}
\put(15,11){\vector(0,-1){1.5}}
\put(13.5,11.1){tungsten}
\put(12,2){\vector(3,2){4}}
\put(10.5,1.4){graphite}
\put(14.5,1){\vector(3,4){3}}
\put(13.5,0.4){LCAL}
\end{picture}
\end{center}
\caption{The mask of the TESLA detector}
\label{fig:mask}
\end{figure}

\section{Calorimetry}
\label{sec:calo}

To measure jet energies the so called ``energy flow'' concept is proposed.
In this scheme the energy is measured as the sum of the charged particle
energies, which contribute about 60\% to a typical hadronic jet, the neutral
electromagnetic energy measured in the electromagnetic calorimeter,
around 30\% of a jet, and the energy of neutral hadrons
measured in the hadronic calorimeter, representing only the remaining 10\% of
the energy. In principle this method allows a very good jet energy resolution,
since the tracking resolution is below a few percent, while the resolution
of a typical hadron calorimeter is around $100\%/\sqrt{E}$. 
However, in this concept one needs a very good spatial resolution of
the calorimeters to separate showers from charged and neutral particles.

The setup of the proposed calorimeters can been seen in
figure \ref{fig:fuldet} . In the
barrel and endcap the electromagnetic (ECAL) and hadronic (HCAL) calorimeters,
which are inside the coil are followed by the instrumented iron return yoke
which is used as a tail catcher and for muon identification.
In the very forward region the mask is instrumented for
electromagnetic calorimetry.

For the ECAL two options are under study, a SiW calorimeter and a lead
scintillator sandwich with shashlik readout. The SiW option offers
several advantages. Due to the high density of tungsten one can pack
25 radiation length into 20\,cm thickness. The small Moliere radius of
tungsten ($\sim 1 \, \cm$) together with a $ 1 \, \cm^2$ Si-pad size offers a
superb lateral resolution and with around 40 layers also a very good
longitudinal resolution is possible. However the detector is fairly expensive
($\sim 130$\meuro). For the shashlik option a readout granularity of 
$3 \times 3 \, \cm^2$ has been studied. Longitudinally a granularity of
two is possible using scintillators of different decay times. The
shashlik calorimeter is about a factor six cheaper than for SiW,
but with worse performance which still has to be quantified.

For the hadronic calorimeter again two options have been studied. One
option is a scintillating tile calorimeter with $5 \times 5\, \cm^2$
tiles in the front part and $25 \times 25 \, \cm^2$ tiles in the back
part. The granularity is limited by the possibility to get the fibres
out of the detector. As an alternative a so called ``digital
calorimeter'' has been proposed. If the granularity of the calorimeter
is high enough the energy resolution using only the number of hit
cells is better than by using the total amplitude.
Following this observation as a second possibility a ``digital
calorimeter'' has been proposed, consisting of $1 \times 1 \, \cm^2$
Geiger counters which are read out binary. The very high granularity
is well matched to the SiW-ECAL allowing an almost perfect separation
of showers from charged and neutral particles.

The detailed setup of the mask is shown in figure \ref{fig:mask}. The
two parts labelled ``LAT'' and ``LCAL'' are equipped as
calorimeters. The LAT is reasonably clean so that it can be used at
least to veto two-photon events. The LCAL receives much
background from $\ee$-pairs from beamstrahlung. Probably it is only usable
for a fast luminosity measurement for machine tuning.

\section{Detector performance}
\label{sec:perf}
\begin{figure}[htb]
\begin{center}
\includegraphics[width=0.4\linewidth,bb=20 3 525 515]{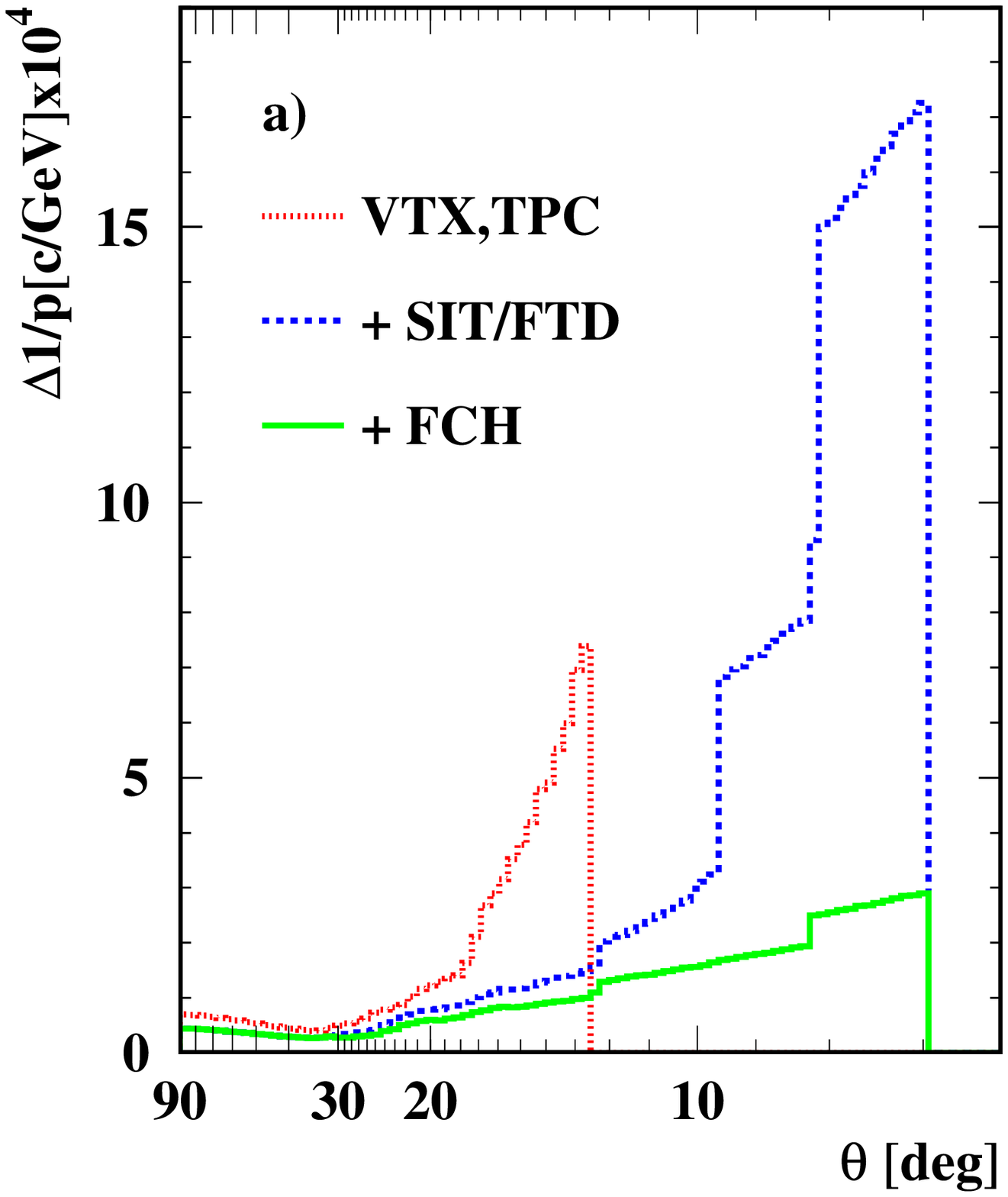}
\includegraphics[width=0.4\linewidth,bb=20 3 525 515]{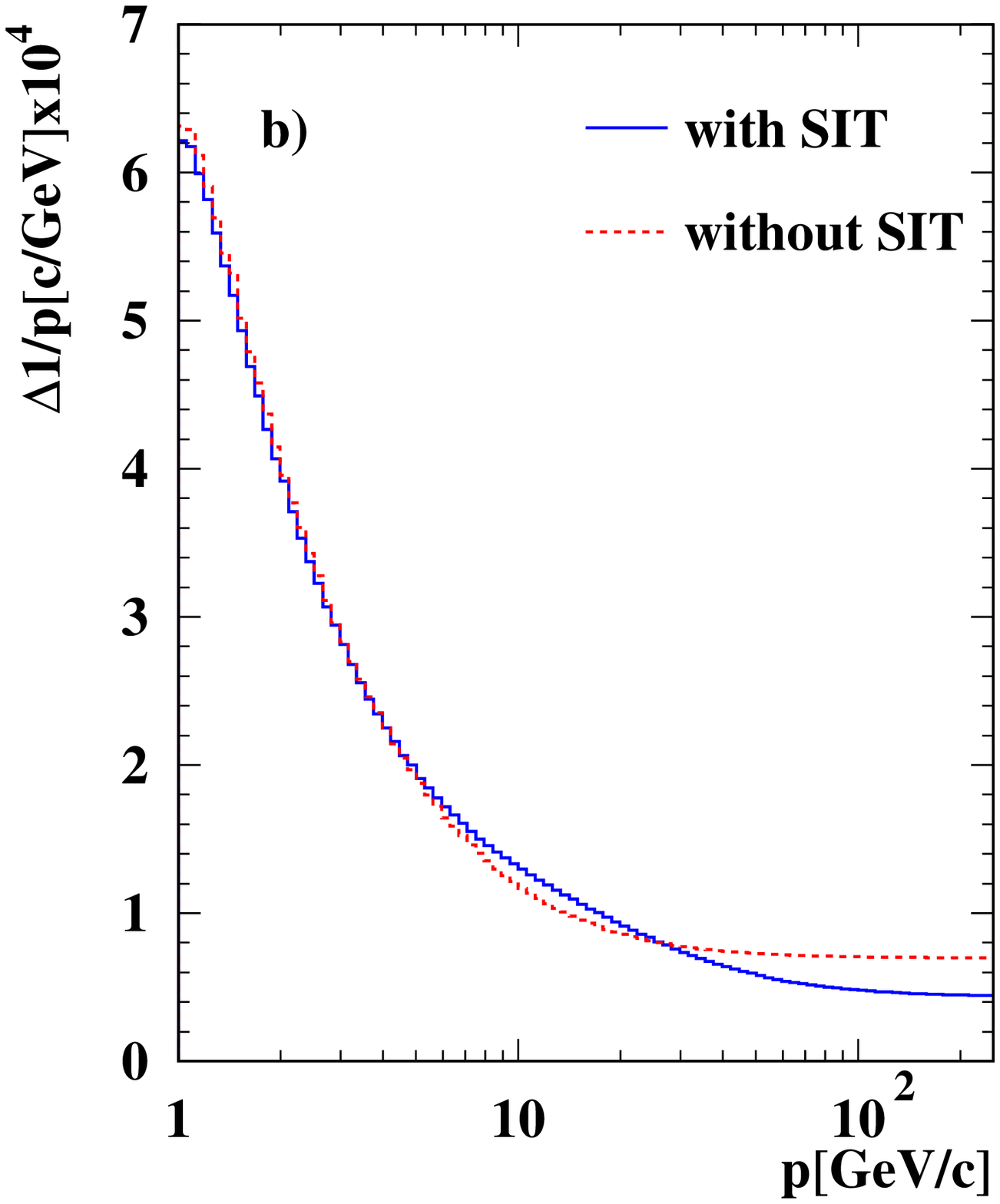}
\end{center}
\caption{Momentum resolution of the tracking system, a) for 250\,GeV
  muons as a function and b) as a function of the momentum for 
  $\theta = 90^\circ$.}
\label{fig:mres}
\end{figure}

Figure \ref{fig:mres} shows the momentum resolution for $250 \GeV$
muons as a function of the polar angle and for particles at $\theta = 90^\circ$
as a function of the momentum.
Over the full tracking region a unique charge
identification is possible up to highest momenta and in the central
region $\Delta \frac{1}{p} \approx 4 \cdot 10^{-5} /\GeV$ is reached. With
this resolution the Z- and Z-recoil mass resolution in 
$\ee \rightarrow {\rm ZH},\, {\rm Z} \rightarrow \mumu$ is dominated
by the Z-width and the beam energy spread.

With the CCD vertex detector the impact parameter resolution is given by
$\sigma = 2.9 \oplus 3.9/(p \sin^{3/2} \theta) \micron$. For b- and
c-tagging the SLD package ZVTOP \cite{ref:ZVTOP} has been
used. As detailed in \cite{ref:stefania} this allows in Z decays at rest to 
tag b-quarks with 70\% efficiency at 95\% purity and to tag c-quark
with 40\% efficiency at 85\% purity.

\begin{figure}[htb]
\begin{center}
\includegraphics[width=0.7\linewidth]{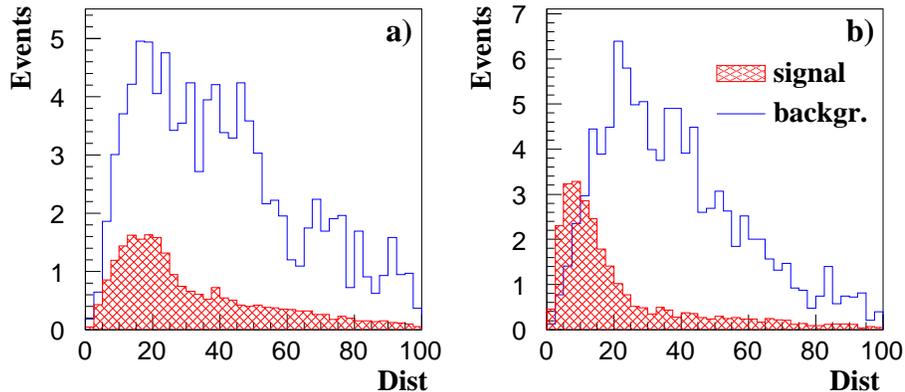}
%\vspace{-0.5cm}
\end{center}
\caption{Mass-distance variable
    for $\ee \rightarrow {\rm ZHH}$ and background assuming
    a): $\Delta E/E = 60\%(1+|\cos\theta_{\rm{jet}}|)/\sqrt{E}$ or
    b): $\Delta E/E = 30\%/\sqrt{E}$.}
\label{fig:eflow}
\end{figure}

An energy flow resolution of $\Delta E/\sqrt{E} = 30\%$ had been reached 
with the SiW-ECAL and the digital HCAL \cite{ref:eflow}. 
As can be seen from figure \ref{fig:eflow} this resolution is needed for
example to measure the Higgs self coupling from 
$\ee \rightarrow {\rm ZHH}$. A comparable analysis for the shashlik ECAL
does not exist yet.

\section{conclusions}
\label{sec:conc}

For the TESLA collider a detector has been designed with which the
proposed physics from $\sqrt{s}=91\GeV$ to $\sqrt{s}=800\GeV$ can be 
well measured.
The degradation of the physics signals due to the finite 
resolution is minimal. The necessary R\&D for the proposed detector
is explained in other contributions to this session.
The total cost of the detector has been estimated to be between 160\,\meuro{}
and 280\meuro{} depending on the calorimeter choice.

% If you have acknowledgements, this puts in the proper section head.
\begin{acknowledgments}
I would like to thank the members of the TESLA detector group for their work 
on the detector concept, the pleasant atmosphere in the group and for their 
help preparing the talk and the manuscript.
\end{acknowledgments}

%\bibliography{e3_moenig_029}

\end{document}